# Interconnection of point defect parameters in solids with bulk properties: A review

**P. Varotsos**

Solid State Section, Physics Department, University of Athens, Panepisthmiopolis, Zografos 157 84, Athens, Greece

E-mail: pvaro@otenet.gr

**Abstract**

Two models have been proposed for the interconnection of the defect Gibbs energy $g^i$ with bulk properties almost 60 and 30 years ago, respectively. The one, proposed by Zener, assumes that $g^i$ can be accounted for the work that goes into straining the lattice and hence it is proportional to the shear modulus of the solid. The other, considers that, since $g^i$ corresponds to an isobaric and isothermal process, it should be proportional to the isothermal bulk modulus and the mean volume per atom. The results of these two models are compared for different processes (defect formation, self-diffusion activation, hetero-diffusion) in a variety of solids including metals (fcc, bcc and tetragonal) as well as solids that exhibit superionic behavior. We find that the latter model does better than the former.

## I. INTRODUCTION

The interrelation between the defect Gibbs energy $g^i$ (where the superscript i stands for the corresponding process, i.e., defect formation, f, migration, m, or self-diffusion activation, act) and bulk properties in solids is a long standing problem. Almost 60 years ago, Zener[1-3] (for a review see Ref. 4) proposed his celebrated model according to which $g^i$ can be expressed through the shear modulus, $\mu$, while 30 years later[5-7] another model -hereafter called $cB\Omega$ model- argued (e.g., see Ref. 8) that $g^i$ should be proportional to the isothermal bulk modulus B and the mean volume $\Omega$ per atom. The interest on these models that interrelate point defect parameters with macroscopic elastic properties (and this is why they are usually termed "elastic" models), has been recently renewed in view of the following findings in diverse fields: First, the latter model was found to be consistent with the parameters that describe the time-dependent polarization arising when changing the rate of uniaxial stress in an ionic crystal or by the indenter penetration into its surface. This plays an important role in clarifying the generation mechanism of transient electric signals that are observed[9-12] before earthquakes. Second, in high Tc-superconductors, and in particular when doping $YBa_2Cu_3O_{7-\delta}$ with alkaline earth elements, it was found[13] that the formation energy for a Schottky defect is compatible with the expectations of the cBΩ-model. Third, these "elastic" models seem to provide a challenging basis for explaining the non-Arrhenius temperature dependence of the viscosity of the glass forming liquids upon approaching the glass transition[14-16], whose full understanding –in spite of significant advances during the last decade (e.g., see Ref. 17)- still remains a mystery. It is, however, not yet clear which of the aforementioned two models enables a better explanation of the data. It is the aim of the present paper to further evaluate these models, mainly focusing on defect formation and self-diffusion activation parameters.

**A. Point defect parameters. Background**

The defect formation parameters are defined, as explained in detail in Ref. 8, by comparing a real (i.e., containing defects) crystal either to an *isobaric* ideal (i.e., *not* containing defects) crystal **or** to an *isochoric* ideal crystal. Thus, there are two different families of defect formation parameters, which are interconnected through thermodynamic relations[8]. The isobaric parameters are defined in terms of the corresponding Gibbs energy ($g^f$) as follows (cf. P, T stand for the pressure and temperature, respectively):

$$s^f = -\frac{dg^f}{dT}|_P \qquad (1)$$

$$h^f = g^f - T\frac{dg^f}{dT}|_P, \text{ and hence } h^f = g^f + Ts^f \qquad (2)$$

$$\upsilon^f = \frac{dg^f}{dP}|_T \qquad (3)$$

where $s^f$, $h^f$ and $\upsilon^f$ designate the defect formation entropy, enthalpy and volume, respectively.

When a single mechanism is operating in mono-atomic crystals, the self-diffusion activation process is described in terms of the activation Gibbs energy $g^{act}$, which is the sum of the Gibbs energies $g^f$ and $g^m$ related to the formation and migration processes, respectively. The activation entropy $s^{act}$ and the activation enthalpy $h^{act}$ are then defined accordingly (e.g., see Ref. 8):

$$s^{act} = -\frac{dg^{act}}{dT}\Big|_P \quad \text{and} \quad h^{act} = g^{act} + Ts^{act}$$

and the diffusion coefficient D is given by

$$D = \gamma \alpha^2 \nu \exp(-\frac{g^{act}}{k_B T}) \qquad (4)$$

where $\gamma$ is a numerical constant depending on the diffusion mechanism and the structure, $\alpha$ stands for the lattice constant, and $\nu$ the attempt frequency which is of the order of the Debye frequency $\nu_D$. In most cases the plots $ln$D versus 1/T (for P=const) and $ln$D versus P (for T=const) are found to be straight lines which reflects that $g^{act}$ varies linearly upon increasing the temperature (for P=const.) and pressure (for T=const.), respectively. There are cases, however, e.g., Na, in which *both* these plots are found to deviate strongly from linearity.

**B. The model proposed by Zener**

Following the wording in Refs. 1, 2, when studying diffusion process, Zener proposed the following: If all the work $g^i$ went into straining the lattice, we would anticipate that:

$$g^i \propto \mu \qquad (5)$$

$$d(\frac{g^i}{g_0})/dT = -d(\frac{\mu}{\mu_0})/dT \qquad (6)$$

where µ refers to an appropriate elastic modulus, and the suffix "0" refers to the absolute zero temperature, or perhaps more correctly, to the extrapolation of the low temperature values to the absolute zero temperature. Since not all of $g^i$ goes into straining the lattice, we anticipate

that the left member of Eq. (6) may be somewhat smaller than the right member. We therefore obtain:

$$\frac{s^i}{h^i} = -\lambda d(\frac{\mu}{\mu_0})/dT \qquad (7)$$

where $\lambda$ is a numerical coefficient less than, but of the order of unity. Zener[1] proceeded to an application of Eq. (7) to various metals by putting it into the form:

$$\frac{s^i}{h^i} = -\frac{\lambda}{T_m} d(\frac{\mu}{\mu_0})/d(\frac{T}{T_m})$$

-where $T_m$ is the melting temperature- by using as $\mu$ the shear modulus (cf. where it was not experimentally available, Young's modulus was used) and calculated that the most recent data at the time were consistent with Eq. (7) if $\lambda=0.55$ and 1 for fcc and bcc metals, respectively. By inserting Eq. (5) into Eq. (3) we get:

$$\frac{\upsilon^i}{g^i} \approx \frac{1}{\mu}\frac{d\mu}{dP} \qquad (8)$$

which is the most frequent form through which Zener's seminal model is checked when performing experiments at various pressures.

**C. The cBΩ model**

We now summarize the alternative model, i.e., the cBΩ model, in which the defect Gibbs energy $g^i$ is interconnected with the bulk properties of the solid through the relation[5-7] (cf. for its justification see Ref. 8):

$$g^i = c^i B\Omega \qquad (9)$$

where $c^i$ is dimensionless which -to the first approximation- can be considered as independent of temperature and pressure (cf. this approximation holds if the ratio $\Delta P/B$ -where $\Delta P$ refers to the pressure range under consideration- is appreciably smaller than unity; otherwise a correction factor should be also taken into account, see p. 126 of Ref. 8). The superscript i in Eq. (9) refers, as mentioned, to the defect processes under consideration, i.e., f, act, m (formation, self-diffusion activation and migration, respectively), and of course $c^f \neq c^m \neq c^{act}$.

By inserting Eq. (9) into Eqs. (1), (2) and (3), we find[5-8,18]

$$s^i = -c^i \Omega (\beta B + \frac{dB}{dT}|_P) \qquad (10)$$

$$h^i = c^i \Omega (B - T\beta B - T\frac{dB}{dT}|_P) \qquad (11)$$

$$\upsilon^i = -c^i \Omega (\frac{dB}{dP}|_T - 1) \qquad (12)$$

where β is the thermal (volume) expansion coefficient. These equations reflect that the ratios $s^i/h^i$, $\upsilon^i/h^i$ and $\upsilon^i/g^i$ depend *solely* on bulk properties, i.e.,

$$\frac{s^i}{h^i} = -\frac{\beta B + \frac{dB}{dT}|_P}{B - T\beta B - T\frac{dB}{dT}|_P} \qquad (13)$$

$$\frac{\upsilon^i}{h^i} = \frac{\frac{dB}{dP}|_T - 1}{B - T\beta B - T\frac{dB}{dT}|_P} \qquad (14)$$

$$\frac{\upsilon^i}{g^i} = \frac{1}{B}(\frac{dB}{dP}|_T - 1) \qquad (15)$$

We now explain how D can be determined -on the basis of the cBΩ model- at any temperature and pressure from a single measurement: By inserting Eq. (9) into equation (4) we get

$$D = \gamma \alpha^2 \nu \exp(-\frac{c^{act} B \Omega}{k_B T}) \qquad (16)$$

Let us first focus on the temperature variation of D at constant pressure. If the value $D_1$ has been found experimentally at a temperature $T_1$, the value of $c^{act}$ can be determined because the pre-exponential factor $\gamma \alpha^2 \nu$ is roughly known. Even if an error of a factor of 2 is introduced by setting $\nu$ equal to $\nu_D$, the value of $c^{act}$ remains practically the same[6]. Hence, once the value of $c^{act}$ has been determined from $D_1$, the value of $D_2$ for a temperature $T_2$ can be found through Eq. (16) if the elastic data and the expansivity data are available for this temperature. Since $D_2$ eventually differs by orders of magnitude from $D_1$, the frequency factor $\nu (= \nu_D)$ can be approximately considered as constant[6]. Note also that, once $c^{act}$ has been determined from $D_1$, the values of $s^{act}$, $h^{act}$ and $\upsilon^{act}$ can be directly calculated at any temperature by means of Eqs. (10), (11) and (12), respectively. By the same token the D value can be studied at any pressure (for T=const) if the value of $c^{act}$ is determined -through Eq. (16)- from the value $D_1$ measured at a single pressure $P_1$. Furthermore, we note that the same procedure, either for P=const or T=const, can be also applied to mixed ionic solids since their B-values used in Eq. (9), can be estimated in terms of the corresponding B-values of the end members.

In view of the above, the present paper is organized as follows: In Section II, we investigate the question of whether "c" can be actually considered as constant in the cBΩ model. Along these lines we study whether this approximation enables the description of the temperature variation of self diffusion especially in a solid in which D varies by several orders of magnitude and furthermore it exhibits an upward curved $l$nD vs 1/T plot in the high temperature range. This is the case, for example of Na. Furthermore, in the same Section, the case of carbon diffusing in α-Fe is studied in a wide temperature range in which D varies by almost 15 orders of magnitude. In Section III, we proceed to a direct comparison of the two models when they are applied to the vacancy formation process under pressure in a noble metal like Al and in a tetragonal one, i.e., In, as well as to the self-diffusion process in the latter. The comparison is also extended to the pressure variation of the ionic conductivity of β-AgI. In Section IV, since an improvement of Zener's model has been suggested in the frame of dynamical theory of the defect migration in solids, we apply this improvement to the fluorine vacancy and fluorine interstitial motion in cubic $PbF_2$ as well as to the association parameters of the complexes formed in β-$PbF_2$ when it is doped with monovalent cations, and the relevant results are compared to the predictions of the cBΩ model. In Section V, we present our conclusions.

## II. DOES "c" IN THE cBΩ MODEL REMAIN CONSTANT?

We first present our study on the self-diffusion in Na and then turn to carbon diffusing in α-Fe.

**A. Self-diffusion in Na**

In Na, radiotracer measurements were carried out[19] at ambient pressure from 194.4°K up to the melting point. The most popular explanation for the observed upward curvature in the high temperature region of the $l$nD vs 1/T plot is the following: It results from the superposition of two or more diffusion mechanisms, namely the monovacancy-divacancy model (for a compilation of early references see Ref. 8). An alternative explanation was also forwarded long ago[20], which proposed that this curvature can be described in the frame of a *single* mechanism (i.e., monovacancy) *if* the self-diffusion parameters (enthalpy, entropy and volume) exhibit an appropriate temperature dependence. Here, in the latter frame, following Ref. 8, we shall show that the whole self-diffusion curve of Na can be described in terms of the cBΩ model. In other words, the latter model inherently contains the appropriate temperature dependence of the self-diffusion parameters that can fully account for the observed curvature in the $l$nD vs 1/T plot even if a single defect mechanism is operating. In order to achieve this goal, we use the expansivity and elastic data explained below.

The high temperature expansivity data was taken from Adlhart et al.[21]. Concerning the elastic data, we note that the conversion of the adiabatic bulk modulus $B_S$ to the isothermal one (B) has been done by the standard thermodynamical manner using the specific heat data reported in Table 1 of Martin[22]. Unfortunately, there is no single elasticity measurements covering the whole temperature range, i.e., from 194.4°K up to the melting point, in which -as mentioned above- diffusion data have been reported by Mundy[19]. Thus, we face the difficulty to select low temperature data (195°K up to room temperature, RT) and to join them with data obtained above RT. Along these lines, and in order to lower the systematic errors[8], we take the average of the B values reported for the dry-ice point (195°K) and RT. In particular, at 195°K, the B values reported in Refs. 23, 24 and 25 (as the latter one was corrected in Ref. 26) are ~7.0, 6.5 and 6.72 GPa, which result in an average value 6.74 GPa. Similarly, at 299°K the resulting average B value is 6.294 GPa when using the values 6.41, 6.3, 6.16 and 6.305 GPa from the data of Refs. 23, 26, 27 and 28, respectively. Note that the latter average B value, which was also used in Ref. 8 (see p. 199), differs only slightly from the value of 6.310 (80) GPa obtained recently by Hanfland et al.[29] by using high-resolution angle-dispersive synchrotron x-ray diffraction. Taking now the linear interpolation of these two average values at 195 and 299°K, we obtain the B values of the second column of Table 1 from 194.4 up to 313°K. Above this temperature the values of this column come from a least squares fit to a straight line of the B values measured by Fritsch et al.[30].

We now check whether the whole measured[19] diffusion curve can be reproduced by the cBΩ model with a constant $c^{act}$ value. The latter is determined by using the experimental value[19] $D_1=1.59\times10^{-12}$cm$^2$/s at the lowest temperature $T_1=194.4$°K of the measurements at which $\nu_D=2.9\times10^{12}$s$^{-1}$ (see Ref. 31) and $\alpha=4.255$ Å (see Ref. 24, and hence $\Omega=\alpha^3/2=38.517\times10^{-24}$cm$^3$). By inserting these values (as well as $\gamma=0.727$ for the monovacancy mechanism) into Eq. (16) for T=194.4°, we find $c^{act}=0.223$ when also using the value of B= 6.743 GPa at that temperature (see also the second column of Table I). Since $c^{act}$ is now known, the D values can be calculated -by means of Eq. (16)- at various temperatures up to the melting point. The results of this calculation are inserted with crosses in Fig. 1 at those temperatures at which experimental D values have been reported by Mundy[19] (the latter are shown by open circles). For the reader's convenience, at some of these temperatures, the numerical values are also given in Table I along with the uncertainty resulting from a plausible experimental error of ~2% in the BΩ value used for each temperature. An inspection of Fig. 1 reveals a striking agreement between the calculated and the experimental D values, which can be alternatively judged from their percentage deviation given in the last column (by considering also that the experimental errors of the D values reported by Mundy[19] lie between 1 and 3%). In other words, in spite of the fact that the *l*nD vs 1/T exhibits a clear upward curvature and that the D values vary by ~5 orders of magnitude (see

Fig. 1), the data can be described on the basis of Eq. (16) without the necessity of using *any* adjustable parameters.

**B. Carbon diffusing in α-Fe**

The experimental data[32,33] of carbon diffusing in α-Fe show that, in the temperature range from 233.9 to 1058°K, the value of D increases by more than 15 orders of magnitude. The $B_S$ values have been measured in the region 298 to 1173°K by Dever[34] and we convert them to the corresponding B ones -in the standard thermodynamic manner, as above- by using the expansivity and specific heat data given in the literature[35]. The determination of $c^{act}$ is now made at the lowest temperature $T_1$=234°K at which D has been measured. For this temperature, the values $D_1$=5.70×10$^{-21}$ or 5.56×10$^{-21}$ cm$^2$/s have been reported in Refs. 32 and 33, respectively, whereas the expansivity and elastic data mentioned above indicate that $\Omega$=11.75×10$^{-24}$cm$^3$ and B=167.2 GPa. Finally, we assume that the diffusion proceeds by interstitials in octahedral sites and hence we set[34] $\gamma$=1/6. As for the attempt frequency, $\nu$, we consider that for a given matrix and mechanism, it depends roughly on the mass of the diffusant and, in absence of a better information, we rely on the usual approximation

$$\frac{\nu}{\nu_D} = (\frac{m_m}{m_j})^{1/2} \qquad (17)$$

where $m_m$, $m_j$ denote the mass of the matrix (m) and the diffusant (j), respectively and $\nu_D$ is ~9.8×10$^{12}$s$^{-1}$ (see Ref. 31). By inserting these values into Eq. (16), we find

$$c^{act} = 0.06697$$

Note that the value of $\nu$ through the approximate relation (17) may be in error by a large factor, which reflects of course an error in the determination of $c^{act}$. In the present case, it was checked that a plausible inaccuracy of $\nu$ by a factor of 5 results in an error of ~4% in the value of $c^{act}$ mentioned above. By inserting the above value of $c^{act}$ into Eq. (16) and using the appropriate data of B and $\Omega$, we compute D for every temperature. For example, at the highest temperature T=1058°K -by using the values B=127.5GPa and $\Omega$=12.18×10$^{-24}$cm$^3$- we compute D=2.13×10$^{-6}$cm$^2$/s, which is in satisfactory agreement with the experimental value

D=1.76×10⁻⁶cm²/s or D=2.11×10⁻⁶cm²/s from Refs. 32 and 33, respectively. Recall that this agreement is achieved although the D value at T=1058°K exceeds that at T=234°K by 15 orders of magnitude. This evidently means that the parameter $c^{act}$ in the cBΩ model does not exhibit any significant change despite of a considerable temperature increase. For the sake of comparison, we remark that Dever[34] (see his Fig. 5) employed the Zener model, but could *not* quantitatively reproduce this diffusion plot.

**III. COMPARING cBΩ- WITH THE ZENER-MODEL**

**A. Vacancy formation in Al**

In order to make the comparison of the two models for the vacancy formation process, we intentionally select Al (for the reason explained below) which is one of the best studied noble metals with differential dilatometry techniques[36-38] that give the directly measured value of the vacancy concentration:

$$\frac{n}{N} = 3(\frac{\Delta L}{L_0} - \frac{\Delta \alpha}{\alpha_0})$$

from simultaneous measurements of the relative change in the macroscopic length L and in the lattice parameter α. The most recent results of these techniques[38] (with somewhat improved accuracy, i.e., to some parts in 10⁶) show that -among the noble metals Al, Ag, Au and Cu studied- only for Al clear indications exist that the Arrhenius plot $ln(n/N)$ vs 1/T is curved upwards when comparing high temperature dilatometry data to low temperature positron and electrical resistivity (after quenching)[39,40] measurements. This curvature makes Al as one of the most proper materials in order to check the capability of an elastic model to describe the temperature dependence of the vacancy concentration.

The isothermal bulk modulus in Al has been measured by Tallon and Wolfenden[41] from RT to just 20°K below the melting point. As for the Ω values, we consider the lattice constant 4.0493 21Å reported in Ref. 42 for 21°C (which practically coincides with the value 4.05109 Å reported in Ref. 37) and then use the measured values of $\Delta L/L_0$ that have been expressed in a form of a polynomial by Guerard et al.[37]. We now determine the value of $c^f$ from the relevant measurements at T=913°K, which is the highest temperature at which B has been reported[41], i.e., B=56.7GPa. For this temperature, the value resulting from the $\Delta L/L_0$ polynomial of Guerard et al.[37] is n/N=8.5776×10$^{-4}$ (and $\Omega$=17.5353Å$^3$). Alternatively, if ones uses the effective values of h$^f$ and s$^f$ they report[37], one finds from the relation n/N=exp(-g$^f$/k$_B$T) the value n/N=7.6829×10$^{-4}$. By inserting the former n/N-value into the cB$\Omega$ expression (see Eq. (9) for i=f):

$$\frac{n}{N} = \exp(-\frac{c^f B\Omega}{k_B T}) \qquad (18)$$

we find c$^f$=0.08948, while the latter n/N value leads to a somewhat larger value, i.e., c$^f$=0.090877. Once the c$^f$ value has been determined, the n/N value can be calculated at any temperature by inserting the corresponding values of B and $\Omega$ (see Table 9.2 of Ref. 8) into Eq. (18).

Let us now calculate, for example, the n/N value at the lowest temperature T=553°K at which the resistivity ($\rho$) measurements[39,40] on quenched samples can be considered as giving reliable[43] n/N values (using $\rho$=1$\mu\Omega$ cm for atomic percent of vacancies). At this temperature, we have B=67.3GPa and $\Omega$=16.9311 Å$^3$ which, when inserted into Eq. (18), lead to $\frac{n}{N} = 1.58^{+0.34}_{-0.29} \times 10^{-6}$ (cf. the uncertainty comes from a plausible uncertainty of 1.5% in the c$^f$ value for the reasons explained above), which is in good agreement with the experimental results[39,40]. This n/N value, which differs by almost ~3 orders of magnitude from the one at T=913°K, mainly resulted from the temperature variation of B from 56.7 to 67.3GPa.

We now compare the above result with that obtained from the Zener model. We consider that the two independent pure shear constants of a cubic crystal are: C$_{44}$ and C′=(C$_{11}$-C$_{12}$)/2 (cf. *no* isothermal-adiabatic distinction exists for the shear constants). Hence, we shall investigate here whether the experimental n/N data can be well described if we insert in Zener's expression either $\mu$=C$_{44}$ or $\mu$=C′. If we use, instead of B in Eq. (18), the values[41] ($\mu$)=C$_{44}$=17.8 and 24.0GPa for T=913 and 553°K, respectively, we find that the vacancy concentration decreases from the experimental value $\approx 8.6\times 10^{-4}$ at T=913°K to a predicted one $\approx 1.48\times 10^{-7}$ at T=553°K. The latter concentration differs from the experimental one (at T=553°K) as well as from that calculated by the cBΩ model by almost one order of magnitude. Alternatively, if one uses the quantity $(\mu)=C'=(C_{11}-C_{12})/2$, the values of which are[41] 12.2 and 18.9GPa for T=913 and 553°K, respectively, we find that the measured vacancy concentration of $8.6\times 10^{-4}$ (at T=913°K) should decrease down to $1.44\times 10^{-8}$ at T=553°K. The latter is, however, $10^2$ times too small compared to either the experimental value or the one obtained from the cBΩ model.

In summary, if one applies Zener's model and hence uses the shear modulus μ, namely either $C_{44}$ or $(C_{11}-C_{12})/2$, the predicted temperature variation of the vacancy concentration in Al deviates strongly from the experimental behavior, while the latter is in full accordance to the cBΩ model.

**B. Pressure dependence of vacancy formation and self-diffusion in In**

We now proceed to an example of a tetragonal metal, i.e., In, and study the pressure variation of both the vacancy formation and the self diffusion activation processes.

Let us start from the elastic data. The pressure derivatives of the elastic constants have been measured by Flower et al.[44]. Concerning the adiabatic bulk modulus $B_S$ and its pressure derivative, the following values have been reported for T=298°K:

$$B_S=43.1 \text{ GPa}$$

$$\frac{dB_S}{dP}|_T \cong 6.68$$

Since the derivatives $(dB_S/dP)_T$ and $(dB/dP)_T$, for T=298°K, differ by no more than a few percent, we adopt here the approximation $(dB_S/dP)_T \approx (dB/dP)_T$. As for the conversion of the adiabatic $B_S$ value to B, we use the thermodynamic relation:

$$B_S=B(1+\gamma_{th}\beta T) \qquad (19)$$

where $\gamma_{th}$ stands for the thermal Gruneisen parameter. Values of $\gamma_{th}(\equiv \beta V/B_S C_P)$ have been given as[45] 2.5 from RT down to temperatures of the order of $\Theta_D/4$ and 2.419 in Ref. 46 at RT.

Since the former value is in better agreement with the mean acoustic Gruneisen parameter $\gamma_H$=2.56 found by Flower et al.[44], we use here this value, i.e., $\gamma_{th}$=2.5. By applying Eq. (19) and using the value[47] $\beta$=9.63×10$^{-5}$K$^{-1}$, we find B=40.2GPa (for T=298°K). Hence, the aforementioned elastic data reveal:

$$\frac{1}{B}(\frac{dB}{dP}|_T - 1) = 14.1 \times 10^{-11} m^3/J \qquad (20)$$

Concerning the shear modulus and its pressure derivative, Flower et al.[44] reported the following values (for T=298°K):

$$\mu = 1/2 \ (C_{11}-C_{12}) = 2.6 \text{GPa}$$

$$d[1/2 \ (C_{11}-C_{12})] \ / \ dP \approx 0.56$$

which lead to:

$$\frac{1}{\mu}\frac{d\mu}{dP} = 21.6 \times 10^{-11} m^3/J \qquad (21)$$

The relations (20) and (21), if we recall Eqs. (15) and (8), provide the expected values for $\upsilon^i/g^i$ according to the cBΩ and the Zener's model, respectively,.

Let us now investigate the vacancy formation parameters in In. The experimental data of Ref. 48 give $s^f \approx 5k$ and $h^f$=0.54±0.03eV and hence the value of $g^f$, for T=298°K, is $g^f$=0.416eV with a plausible uncertainty of at least ±0.03eV. Furthermore, since the angular correlation measurements under pressure of Ref. 49 give $\upsilon^f$=6.1±0.2 cm$^3$/mol, the experimental value of $\upsilon^f/g^f$ is:

$$\frac{\upsilon^f}{g^f} = 15.2^{+1.7}_{-1.5} \times 10^{-11} m^3/J$$

which is closer to the aforementioned value expected from the cBΩ model (see Eq. (20)) compared to that from the Zener model.

We now proceed to the experimental results of the self-diffusion activation parameters[50] in In. The activation enthalpy $h^{act}$ from the temperature variation of D is 0.81±0.015eV while the activation volume $\upsilon^{act}$=8.1±0.4cm$^3$/mol. In order to estimate the self-diffusion activation entropy, we rely on the value[48] $v\exp(s^m/k_B)=1.2\times10^{13}s^{-1}$ from which -when assuming $\nu \approx \nu_D$ and considering that[31] $\nu_D=1.8\times10^{12}s^{-1}$- we find $s^m=1.9$ $k_B$. This value, when recalling that[48] $s^f=5k$ and $s^{act}=s^f+s^m$, gives $s^{act}=6.9k_B$. Hence, the value of $g^{act}(=h^{act}-Ts^{act})$ for T=298°K is found to be $g^{act}=0.633eV$ with a plausible error of around ±0.015eV. Thus, the experimental value of $\upsilon^{act}/g^{act}$ for the self-diffusion process is:

$$\frac{\upsilon^{act}}{g^{act}}=13.3^{+1.0}_{-1.0}\times10^{-11}m^3/J$$

which is comparable with the value predicted from the cBΩ-model (see Eq. (20)), but *not* with that from Zener's model (see Eq. (21)).

### C. Pressure dependence of the ionic conductivity of β-AgI

We finally turn to a striking example, in which the ionic conductivity σ *increases* upon compression, i.e., the corresponding activation volume $\upsilon^{act}$ is negative. This is the case of β-AgI (B-4 phase)[51], which is in sharp contrast to the behavior found in the usual ionic materials, e.g., alkali halides, in which σ *decreases* upon compression, i.e., $\upsilon^{act}$ is positive (see Ref. 8 and references therein). In particular, the measurements of Allen and Lazarus[51] show that $\upsilon^{act}$ is about -3cm³/mole at RT and becomes more negative as the temperature increase, reaching a value $\upsilon^{act}\approx-9.5cm^3$/mole at T=400°K.

Let us now consider the elastic behavior of this material which has been studied by Shaw[52]. It also exhibits an unusual temperature dependence, i.e., the measurements at RT show that dB/dP has a very small positive value which becomes distinctly negative when the temperature increases. In particular for T≈395°K these measurements show that dB/dP is around -2.

We first apply Eq. (15) at RT. Using the values[52] B≈24.0GPa, (dB/dP)≈0 as well as the value[51] $h^{act}\approx0.8eV$ and taking the approximation $g^{act}\approx h^{act}$, Eq. (15) leads to $\upsilon^{act}=-3cm^3$/mole which is in excellent agreement with the aforementioned experimental value[51] $\upsilon^{act}=-3cm^3$/mole. Repeating the same calculation at 395°K but now considering (dB/dP)=-2, Eq.

(15) gives $\upsilon^{act}$=-9.8cm$^3$/mole which again agrees with the experimental result[51] $\upsilon^{act}$=-9.3cm$^3$/mole.

Let us now focus on temperatures higher than ~400°K, where AgI transforms into the B-23 phase in which a superionic behavior has been observed. In this phase, $\upsilon^{act}$ is experimentally found to have a small positive value, which conforms to the cBΩ model, i.e., Eq. (15), in view of the following: First, the activation enthalpy h$^{act}$ is[51] around a few tenths of eV (thus being markedly smaller than the aforementioned value ~0.8eV at the B-4 phase) and, second, the elastic measurements[52] show that dB/dP is definitely positive having a low but not unreasonable value (with dB/dP>1)[52]. On the other hand, in this phase, the shear modulus (μ) exhibits a definite negative value[52], thus implying -according to the Zener model, see Eq. (8)- a negative activation volume which sharply contradicts the experimental results.

## IV. COMPARING cBΩ- WITH THE ZENER-MODEL AS MODIFIED BY THE DYNAMICAL THEORY

Flynn[53], who developed the dynamical theory of migration in solids, suggested that in the Debye approximation the shear modulus μ in Zener's model has to be replaced by another shear modulus C* given by (see p. 332 of Ref. 53):

$$\frac{15}{2C^*} = \frac{3}{C_{11}} + \frac{2}{C_{11}-C_{12}} + \frac{1}{C_{44}} \qquad (22)$$

and hence, within the dynamical theory, the migration volume $\upsilon^m$ is connected to the migration Gibbs energy g$^m$ through the relation:

$$\frac{\upsilon^m}{g^m} = -\frac{1}{B} + \frac{1}{C^*}\frac{dC^*}{dP} \qquad (23)$$

which obviously differs from the corresponding relation -see Eq. (15) for i=m- of the cBΩ model.

### A. Migration of fluorine vacancy and fluorine interstitial in β-PbF$_2$

Let us now compare the results of Eqs. (15) and (23) for the well studied case of cubic (fluorite structured) PbF$_2$. The adiabatic elastic data reported in Ref. 54, for T=296°K, lead to $B_S$=63.0GPa, $dB_S/dP$=7.13 and $dC^*/dP$=2.56. Here -for the sake of convenience- we approximate that the values of $B_S$ and $dB_S/dP$ are equal to their isothermal ones. Applying Flynn's formula, i.e., Eq. (23), we find:

$$\frac{\upsilon^m}{g^m} = 2.5 \times 10^{-11} m^3/J \qquad (24)$$

while the cBΩ model, through Eq. (15), predicts an appreciably larger value:

$$\frac{\upsilon^m}{g^m} = 9.7 \times 10^{-11} m^3/J \qquad (25)$$

We now turn to the experimental results. For T=300°K, the parameters $h^m$ and $s^m$ given in Table 4 of Ref. 55 lead to $g^m$=0.20 and 0.346V for the fluorine vacancy and interstitial motion, respectively. These values, when also considering the $\upsilon^m$-values reported in Ref. 56 (i.e., 1.9 and 3.5 cm$^3$/mole, respectively) from conductivity measurements under pressure, lead to the following results:

$$\frac{\upsilon^m}{g^m} = 9.9 \times 10^{-11} m^3/J \qquad \text{for fluorine vacancy, and}$$

$$\frac{\upsilon^m}{g^m} = 10.5 \times 10^{-11} m^3/J \qquad \text{for fluorine interstitial}$$

with a plausible experimental uncertainty of around 10% (see pp. 300-303 of Ref. 8). These experimental results agree with the ratio predicted from the cBΩ model –see Eq. (25)- but differ considerably from that obtained on the basis of the Zener model as modified by Flynn[53], see Eq. (24).

The above study referred to cubic PbF$_2$. But when it transforms into the orthorhombic structure, i.e., α-PbF$_2$, the migration volumes for vacancy fluorine motion and interstitial fluorine motion are found[56] to be roughly equal (i.e., 3.7±0.3 and 3.1±0.3 cm$^3$/mole, respectively) -in contrast to the cubic phase where they differ by a factor of two. This behavior again conforms to the cBΩ model, because in the orthorhombic structure the experimental values of the migration enthalpies for the aforementioned two processes are comparable (0.36±0.02 and 0.41±0.02eV, respectively) and hence -through Eq. (14)- the migration volumes should be also roughly equal.

**B. Association parameters of the complexes formed when doping β-PbF$_2$ with alkali ions**

An additional confirmation of the cBΩ-model comes from the following experimental data in the cubic PbF$_2$: When doping PbF$_2$ with monovalent cations, fluorine vacancies are created for reasons of change compensation. A certain percentage (depending on temperature) of these vacancies is attracted to the impurity ions thus forming electric dipoles (complexes), while the rest are dissociated from the dipoles and constitute freely mobile vacancies. The (re)orientation parameters of these dipoles are determined by dielectric loss measurements [57] (from which the migration volume for the re-orientation process have been reported[57]). In addition, accurate low-temperature dc-conductivity measurements under various pressures have been carried out[57] on PbF$_2$ either pure or doped with various alkali metals. At the lower temperatures of these measurements the dc conductivity is due to carries thermally dissociated from these complexes and hence its temperature and pressure variation is governed by the activation parameters $h^{act}$ and $\upsilon^{act}$ given by:

$$h^{act} = 1/2\, h^a + h^{fm}$$

$$\upsilon^{act} = 1/2\, \upsilon^a + \upsilon^{fm}$$

where $h^a$, $h^{fm}$ and $\upsilon^a$, $\upsilon^{fm}$ stand for the enthalpies and volumes for the association (a) process and the free vacancy motion (fm), respectively. For both processes (i=a, fm), since the entropy $s^i$ has a value around a few $k_B$ at the most, we can assume (for T≈300°K) $Ts^i \ll h^i$ and hence $g^i \approx h^i$. Thus, Eq. (15) can be approximately written as:

$$\frac{\upsilon^i}{h^i} \approx \frac{1}{B}\frac{dB}{dP}\bigg|_T - \frac{1}{B} \qquad (26)$$

This relation reveals that, irrespective of the process and the kind of the impurity, the ratio $\upsilon^i/h^i$ *solely* depends on the bulk properties being equal to $(\frac{dB}{dP}\big|_T - 1)/B$.

In order to check this prediction, the values of $\upsilon^i$ have been plotted versus $h^i$ in Ref. 58 for the association region of the conductivity curve for pure PbF$_2$ and for PbF$_2$ doped with Li,

Na, K and Rb. In this plot, the relevant values for the migration processes of the fluorine interstitial and the fluorine vacancy have been also inserted. All these points ($h^i$, $\upsilon^i$) have been found[58] to lie on a straight line (passing through the origin of the axis) with a slope equal to that predicted from Eq. (26), i.e., $\approx 9.7 \times 10^{-11} m^3/J$ (see Eq. (25)). This value, which is in striking agreement with the cBΩ model, strongly deviates from the prediction of Zener's model, as modified by Flynn, see the value given by Eq. (24) (when recalling $g^m \approx h^m$).

## V. CONCLUSIONS

Examples have been presented here in a large variety of solids, i.e., bcc, fcc and tetragonal metals, as well as in two solids that exhibit the so called superionic behavior, i.e., $PbF_2$ and AgI, for various defect processes, i.e., formation, migration, self-diffusion activation and (dis)association of complexes. (The case of β-$PbF_2$ has been also separately discussed elsewhere[59]). In all these cases, the cBΩ model leads to results that agree better with the experimental data compared to those obtained either from the original model of Zener or from a modification proposed for the latter. The same conclusion has been drawn in an early study[8] to other categories of solids, namely alkali- and silver- halides (in which Schottky and Frenkel defects prevail, respectively) as well as rare gas solids (monovacancies). In other words, in all the classes of solids investigated to date, the cBΩ model is found to be superior to Zener's model as far as the defect processes is concerned.

**References**

[1] C. Zener, J. Appl. Phys. **22**, 372 (1951).

[2] C. Wert, and C. Zener, Phys. Rev. **76**, 1169 (1949).

[3] C. Zener, *Imperfections in Nearly Perfect Crystals*, edited by W. Shockley (Wiley, New York, 1952), p. 289.

[4] J. Philibert, Defect Diff. Forum, **249**, 61 (2006).

[5] M. Lazaridou, C. Varotsos, K. Alexopoulos and P. Varotsos, J. Physics C: Solid State **18**, 3891 (1985).

[6] P. Varotsos, and K. Alexopoulos, J. Physics and Chemistry Solids **42,** 409 (1981).

[7] P. Varotsos, and K Alexopoulos, Phys. Rev. B **21**, 4898 (1980).

[8] P. Varotsos, and K. Alexopoulos, *Thermodynamics of Point Defects and their relation with the bulk properties*, *North Holland* (1986) 474 pages

[9] P. Varotsos, N. Sarlis, and M. Lazaridou, Acta Geophysica Polonica **48**, 141 (2000).

[10] P. Varotsos, K. Eftaxias, M. Lazaridou, K. Nomicos, N. Sarlis, N. Bogris, J. Makris, G. Antonopoulos and J. Kopanas, Acta Geophysica Polonica **44**, 301 (1996).

[11] P.A. Varotsos, N. V. Sarlis and E. S. Skordas, EPL (Europhysics Letters) **99**, 59001 (2012).

[12] P. A. Varotsos, N. V. Sarlis, E. S. Skordas, and M. S. Lazaridou, Tectonophysics **589,** 116 (2013).

[13] H. Su, D.O. Welch, and Winnie Wong-Ng, Phys. Rev. B **70**, 054517 (2004).

[14] J.C. Dyre, Rev. Mod. Phys. **78**, 953 (2006).

[15] J.C. Dyre, AIP Conference Proceedings **832**, 113 (2006).

[16] J.C. Dyre, T. Christensen, and N.B. Olsen, J. Non-Cryst. Solids **532**, 4635 (2006).

[17] J. S. Langer, PHYSICS TODAY, February 2007, p. 8; see also Phys. Rev. E **73**, 041504 (2006).

[18] P. Varotsos, and K. Alexopoulos, J. Phys. Chem. Solids **41**, 443 (1980).

[19] J.N. Mundy, Phys. Rev. B **3** 2431 (1971).

[20] H.M. Gilder, and D. Lazarus, Phys. Rev. B **11** 4916 (1975).

[21] W. Adlhart, G. Fritsch, A. Heidemann, and E. Luescher, Phys. Lett. A **47** 91; **48,** 239 (1974).

[22] D.L. Martin, Phys. Rev. **154**, 571 (1967).

[23] R.H. Martinson, Phys. Rev. **178**, 902 (1969).

[24] M.E. Diederich, and J. Trivisonno, J. Phys. Chem. Solids **27**, 637 (1966).

[25] R.I. Beecroft, and C. A. Swenson, J. Phys. Chem. Solids **18**, 329 (1961).

[26] C.E. Monfort, and C.A. Swenson, J. Phys. Chem. Solids **26**, 291 (1965).

[27] W.B. Daniels, Phys. Rev. **119**, 1246 (1960).

[28] G. Fritsch, F. Geipel, and A. Prasetyo, J. Phys. Chem. Solids **34**, 1961 (1973).

[29] M. Hanfland, I. Loa, K. Syassen, Phys. Rev. B **65,** 184109 (2002).

[30] G. Fritsch, M. Nehmann, P. Korpiun, and E. Luescher, phys. stat. sol. (a) **19,** 555 (1973).

[31] R.C.G. Killean, and E.J. Lisher, J. Phys C: Solid State Phys. **8,** 3510; J. Phys. F: Metal Phys. **5,** 1107 (1975).

[32] A.E. Lord, and D.N. Beshers, Acta Metall. **14**, 1659 (1966).

[33] J.R.G. Silva, and R.B. McLellan, Mater. Sci. Eng. **26**, 83 (1976).

[34] D. J. Dever, J. Appl. Phys. **43**, 3293 (1972).

[35] Metals Handbook: Properties and Selection of Metals, 8th ed., edited by Lydon Taylor (American Society for Metals, Cleveland, 1961), Vol. I.

[36] R.O. Simmons, and R.W. Balluffi, Phys. Rev. **117**, 52 (1960).

[37] R. Guerard, H. Peisl, and R. Zitzmann, Appl. Phys. **3,** 37 (1974).

[38] Th. Hehenkamp, J. Phys. Chem. Solids **55**, 907 (1994).

[39] A.S. Berger, S. T. Ockers, M. K. Chason, and R.W. Siegel, J. Nucl. Mater. **69-70**, 734 (1978).

[40] J. Bass, Phil. Mag. **15**, 717 (1967).

[41] J. L. Tallon, and W. Wolfenden, J. Phys. Chem. Solids, **40**, 831 (1979).

[42] A. J. Cornish, and J. Burke, J. Scient. Instr. **42**, 212 (1955).

[43] R. W. Siegel, J. Nucl. Mater. **69-70**, 117 (1978).

[44] S.C. Flower, G. A. Saunders, and Y. K. Yogurteu, J. Phys. Chem. Solids, **46**, 97 (1985).

[45] N. Madaiah, and G. M. Graham, Can. J. Phys. **42**, 221 (1964).

[46]J. Ramakrishnan, R. Bochler, G. Higgins, and G. C. Kennedy, J. Geophys. Res. **83**, 3535 (1978).

[47]R. K. Kirby, T. A. Hahn, and B. D. Rohtrock, American Institute of Physics Handbook ed. D. E. Gray (New York, McGraw-Hill, 1972) p. 4-125.

[48]W. Weiler, and H. E. Schaefer, J. Phys. F: Met. Phys. **15**, 1651 (1985).

[49]J. E. Dickman, R. N. Jeffery, and D. R. Gustafson, Phys. Rev. B **16**, 3334 (1977).

[50]J. Dickey, Acta Metall. **7**, 350 (1959).

[51]P. C. Allen, and D. Lazarus, Phys. Rev. B **17**, 1913 (1978).

[52]G. H. Shaw, J. Geophys. Res. **83**, 3519 (1978); J. Phys. Chem. Solids **35**, 911 (1974); ibid **41**, 155 (1980).

[53]C. P. Flynn, *Point Defects and Diffusion* (Clarendon Press, Oxford 1972).

[54]D. S. Rimai, and R. J. Sladek, Phys. Rev. B **21**, 843 (1980).

[55]A. Azimi, V. M. Carr, A. V. Chadwick, F. G. Kirkwood, and R. Saghafian, J. Phys. Chem. Solids **45**, 23 (1984).

[56]G. Samara, J. Phys. Chem. Solids **40**, 509 (1979).

[57]D. R. Figueroa, J. J. Fontanella, M. C. Wintersgill, A. V. Chadwick, and C. G. Andeen, J. Phys. C **17**, 4399 (1984).

[58]M. Lazaridou, K. Alexopoulos, and P. Varotsos, Phys. Rev. B **31**, 8273 (1985).

[59]P. Varotsos, Solid State Ionics **179**, 438 (2008).

**Table I.** Calculated and experimental self-diffusion in sodium along with the elastic end expansivity data involved in the cBΩ mode

| T (K) | B (GPa) | Ω ($10^{-4}cm^3$) | $D_{exp}$ ($cm^2/s$) | $D_{calc}$ ($cm^2/s$) | ($D_{calc}$- $D_{exp}$)/ $D_{exp}$ % |
|---|---|---|---|---|---|
| 194.4 | 6.743 | 38.517 | $1.59 \times 10^{-12}$ | $1.59 \times 10^{-12}$ | 0 |
| 288 | 6.341 | 39.34 | $3.23 \times 10^{-9}$ | $3.22^{+1.22}_{-0.78} \times 10^{-9}$ | 0 |
| 298.2 | 6.297 | 39.44 | $5.81 \times 10^{-9}$ | $5.52^{+1.67}_{-1.3} \times 10^{-9}$ | -5 |
| 308 | 6.178 | 39.52 | $1.02 \times 10^{-8}$ | $1.05^{+0.31}_{-0.23} \times 10^{-8}$ | +3 |
| 313 | 6.140 | 39.56 | $1.33 \times 10^{-8}$ | $1.38^{+0.38}_{-0.31} \times 10^{-8}$ | +4 |
| 327.9 | 6.101 | 39.69 | $2.65 \times 10^{-8}$ | $2.54^{+0.66}_{-0.54} \times 10^{-8}$ | -4 |
| 357 | 5.855 | 39.95 | $9.81 \times 10^{-8}$ | $9.74^{+2.24}_{-1.86} \times 10^{-8}$ | -1 |
| 362.6 | 5.808 | 40.00 | $1.21 \times 10^{-7}$ | $1.23^{+0.27}_{-0.23} \times 10^{-7}$ | +2 |
| 364.5 | 5.792 | 40.02 | $1.34 \times 10^{-7}$ | $1.33^{+0.3}_{-0.24} \times 10^{-7}$ | -1 |
| 368.7 | 5.757 | 40.06 | $1.64 \times 10^{-7}$ | $1.58^{+0.34}_{-0.3} \times 10^{-7}$ | -4 |

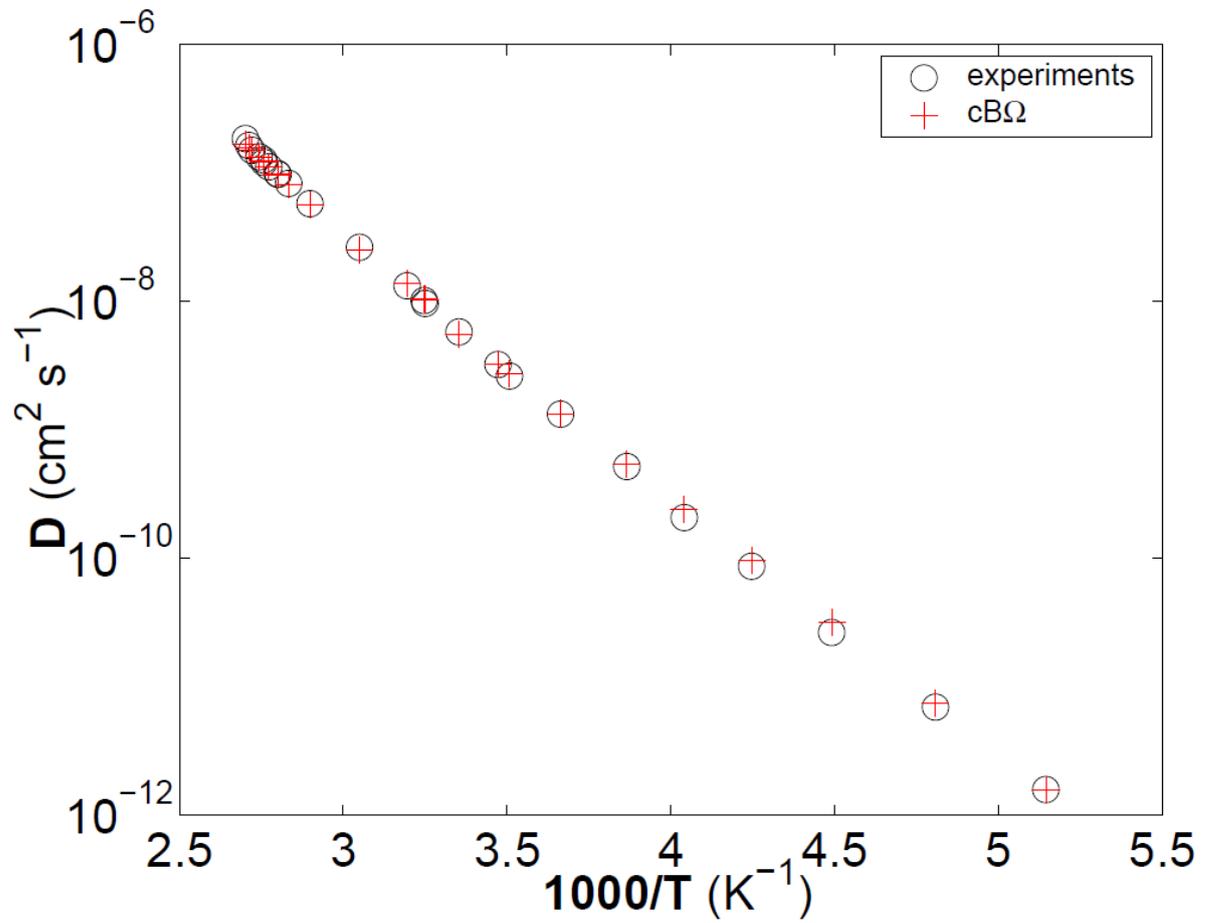

**Fig. 1.** (Color on line) Values of the tracer diffusion coefficient, D, in Na, as measured by Mundy[19] at various temperatures, T, vs 1000/T are shown by the circles. The values calculated using the cBΩ model are shown by the crosses.